\documentclass[article,nojss]{jss}

\usepackage[]{graphicx}\usepackage[]{color}
\makeatletter
\def\maxwidth{ %
  \ifdim\Gin@nat@width>\linewidth
    \linewidth
  \else
    \Gin@nat@width
  \fi
}
\makeatother

\usepackage{Sweave}

\usepackage[latin1]{inputenc}
\usepackage[pdftex]{thumbpdf}
\usepackage[T1]{fontenc}
\usepackage{verbatim}
\usepackage{amssymb,amsmath,amsfonts,amsthm}
\usepackage{graphics}

\title{Calculating Probabilistic Excursion Sets and Related Quantities Using \pkg{excursions}}

\author{
David Bolin\\
Chalmers University of Technology \\
and the University of Gothenburg\And
Finn Lindgren\\
The University of Edinburgh
}

\Plainauthor{David Bolin, Finn Lindgren} 
\Plaintitle{Calculating Probabilistic Excursion Sets and Related Quantities Using Excursions} 

\Abstract{The \proglang{R} software package \pkg{excursions} contains methods for calculating probabilistic excursion sets, contour credible regions, and simultaneous confidence bands for latent Gaussian stochastic processes and fields. It also contains methods for uncertainty quantification of contour maps and computation of Gaussian integrals. This article describes the theoretical and computational methods used in the package. The main functions of the package are introduced and two examples illustrate how the package can be used.}
\Plainkeywords{excursion sets, contour maps, contour curves, simultaneous credible bands, Gaussian integrals, R}

\Address{
  David Bolin\\
  Department of Mathematical Sciences\\
  Chalmers university of technology\\
  S-412 96 Gothenburg, Sweden\\
  E-mail: davidbolin@gmail.com\\
  URL: \url{http://www.math.chalmers.se/~bodavid/}\\
  ~\\
  Finn Lindgren\\
  The University of Edinburgh\\
  James Clerk Maxwell Building\\
  Peter Guthrie Tait Road\\
  Edinburgh EH9 3FD, United Kingdom\\
  E-mail: finn.lindgren@ed.ac.uk\\
  URL: \url{http://www.maths.ed.ac.uk/~flindgre/}
}

\DeclareMathOperator*{\argmax}{arg\,max}

\newcommand{\mapset}{G}

\long\def\symbolfootnote[#1]#2{\begingroup%
\def\thefootnote{\fnsymbol{footnote}}\footnote[#1]{#2}\endgroup}

\theoremstyle{plain} 

\theoremstyle{definition} 

\theoremstyle{remark}

\newcommand{\proper}{\mathsf}

\newcommand{\pN}{\proper{N}}
\newcommand{\mv}[1]{{\boldsymbol{\mathrm{#1}}}}

\newcommand{\trsp}{\ensuremath{\top}}
\newcommand{\md}{\ensuremath{\,\mathrm{d}}}
\newcommand{\exset}[2]{E_{{#1}}^{{#2}}}

\graphicspath{{.}{Rfigs/}}

\IfFileExists{upquote.sty}{\usepackage{upquote}}{}
\begin{document}

\section{Introduction}
The ability to find regions where a stochastic process exceeds a certain level, or is significantly different from some reference level, is important in several areas of application.  Examples in geosciences include studies of air pollution \citep{cameletti12}, temperature \citep{furrer07}, precipitation \citep{sain11}, and vegetation \citep{eklundh,bolin09a}, and similar problems can be found in a wide range of scientific fields including brain imaging \citep{marchini03} and astrophysics \citep{beaky92}. A related problem is uncertainty quantification of contour curves and more generally of contour maps, which are often used to display estimates of continuous surfaces. The number of contours used in a contour map should typically reflect the uncertainty in the estimate, since one should be allowed to draw many contours if the uncertainty of the estimated surface is low and fewer contours if the uncertainty is high. The ability to quantify the uncertainty in the contour map is important if one should be able to choose the number of contours in a rigorous way.

The \proglang{R} \citep{Rteam13} software package \pkg{excursions} \citep{excursions} contains functions for solving these problems for latent Gaussian models (LGMs), which is a large model class that is widely used in applications \citep[see e.g.,][]{rue09}. The computational methods are based on the theory introduced by \cite{bolin12,bolin15contours} and are especially well-suited for models where the latent field has Markov properties. Solving the problems involves compting high-dimensional Gaussian integrals, which can be done more efficiently if Markov properties can be utilized. With the ability to efficiently compute Gaussian integrals, one can also compute simultaneous credible bands for latent Gaussian processes, and more generally for mixtures of Gaussian processes. This was investigated by \cite{bolin2014statistical} and \pkg{excursions} contains a slightly more general implementation of the methods from these papers.

The package supports at least three ways of specifying the model that shuold be analysed. The standard method for purely Gaussian models is to specify the model by providing the parameters of the Gaussian process. For more general models, the input can either be given as Monte Carlo simulations of the process or as the result from an analysis using the \proglang{R}-\pkg{INLA} software package \citep[][package available from \url{http://r-inla.org/download/}]{lindgren2015software}. The package is available to install from the Comprehensive R Archive Network (CRAN) at \url{https://CRAN.R-project.org/package=excursions}. A development version of the package is available via the repository
\url{https://bitbucket.org/davidbolin/excursions}. The development version is updated more frequently, and can be easily be installed directly in \textsf{R} as described on the repository homepage.

The following sections describe the theoretical methods used in the package and provide an introduction to the implementation. Section~\ref{sec:methods}
summarizes the theory, Section~\ref{sec:software} introduces the main functions in the package, and Section~\ref{sec:tutorial} contains two examples that illustrate how the package can be used. Finally, future plans for the package is discussed in Section~\ref{sec:discussion}.

\section{Definitions and computational methodology}\label{sec:methods}

Hierarchical models are of great importance in many areas of statistics. In the simplest form, a hierarchical model has a likelihood distribution $\pi(\mv{Y}|\mv{X}, \mv{\theta})$ for observed data $\mv{Y}$, which is specified conditionally on a latent process of interest, $\mv{X}$,  which has a distribution $\pi(\mv{X}|\mv{\theta})$. For Bayesian hierarchical models, one also specifies prior distributions for the model parameters $\mv{\theta}$. The most important special case of these models are the LGMs, which are obtained by assuming that $\mv{X}|\mv{\theta}$ has a Gaussian distribution. Numerous applications can be studied using models of this form, and these are therefore the main focus of the methods in \pkg{excursions}.

A statistical analysis using an LGM often concludes with reporting the posterior mean $\E(\mv{X}|\mv{Y})$ as a point estimate of the latent field, possibly together with posterior variances as a measure of uncertainty. In many applications, however, reporting posterior means and variances are not enough. As stated in the introduction, one may be interested in computing regions where the latent field exceeds some given threshold, contour curves with their associated uncertainty, or simultaneous confidence bands. In some applications, only a contour map of the posterior mean is reported, where the number of levels in the contour map should represent the uncertainty in the estimate. These are quantities that can be computed with \pkg{excursions} and we now define these in more detail before outlining how they can be computed. For details we refer to \cite{bolin12,bolin15contours}.

\subsection{Definitions}
The main quantities that can be computed using \pkg{excursions} are (1) excursion sets, (2) contour credible regions and level avoiding sets, (3) excursion  functions, (4) contour maps and their quality measures, (5) simultaneous confidence bands. This section defines these in more detail.

Throughout the section, $X(\mv{s})$ will denote a stochastic process defined on some domain of interest, $\Omega$, which we assume is open with a well-defined area $|\Omega|<\infty$. Since it is not necessary for the definitions, we will not explicitly state the dependency on the data, but simply allow $X$ to have some possible non-stationary distribution. In practice, however, the distribution of $X$ will typically be a posterior distribution conditionally on data, $X(\mv{s})|\mv{Y}$. For frequentist models, the distribution of $X$ could also be conditionally on for example a maximum likelihood estimate of the parameters, $X(\mv{s})|\mv{Y},\widehat{\mv{\theta}}$.

\subsubsection{Excursion sets}
An excursion set is a set where the process $X(\mv{s})$ exceeds or goes below some given level of interest, $u$. A where $X(\mv{s})>u$ is referred to as a positive excursion set, whereas a set where $X(\mv{s})<u$ is referred to as a negative excursion set. If $X(\mv{s})=f(\mv{s})$ is a known function, these sets can be computed directly as $A_u^+(f) = \{ \mv{s}\in\Omega; f(\mv{s})>u \}$ and $A_u^-(f) = \{ \mv{s}\in\Omega; f(\mv{s})<u \}$ respectively. If $X(\mv{s})$ is a latent random process, one can only provide a region where it with some (high) probability exceeds the level. More specifically, the positive level $u$ excursion set with probability $\alpha$, $\exset{u,\alpha}{+}(X)$, is defined as the largest set so that with probability $1-\alpha$ the level $u$ is exceeded at all locations in the set,
\begin{equation}\label{def:E}
\exset{u,\alpha}{+} = \argmax_{D}\{|D|:\Prob[D\subset A_u^+(X)]\geq 1 - \alpha\}.
\end{equation}
Similarly, the negative $u$ excursion set with probability $\alpha$, $\exset{u,\alpha}{-}(X)$, is defined as the largest set so that with probability $1-\alpha$ the process is below the level $u$ at all locations in the set. This set is obtained by replacing $A_u^+(X)$ with $A_u^-(X)$ in \eqref{def:E}.

\subsubsection{Contour credible regions and level avoiding sets}
For a function $f$, a contour curve (or set in general) of a level $u$ is defined as the set of all level $u$ crossings. Formally, the level curve is defined as $A_u^c(f) = \left(A_u^+(f)^o\cup A_u^-(f)^o\right)^c$, where $B^o$ denotes the interior of the set $B$ and $B^c$ denotes the complement. Note that $A_u^c(f)$ not only includes the set of locations where $f(\mv{s})=u$, but also all discontinuous level crossings.

For a latent random process $X$, one can only provide a credible region for the contour curve. A level $u$ contour credibility region, $\exset{u,\alpha}{c}(X)$, is defined as the smallest set such that with probability $1-\alpha$ \emph{all} level $u$ crossings of $X$ are in the set. This set can be seen as the complement of the level $u$ contour avoiding set $\exset{u,\alpha}{}(X)$, which is defined as the largest union $M_{u,\alpha}^+ \cup M_{u,\alpha}^-$, where jointly $X(\mv{s})>u$ in $M_{u,\alpha}^+$ and $X(\mv{s})<u$ in $M_{u,\alpha}^-$. Formally,
\begin{equation*}
(M_{u,\alpha}^+(X), M_{u,\alpha}^-(X)) = \argmax_{(D^+,D^-)}\{|D^-\cup D^+| :
\Prob(D^- \subseteq A_u^-(X),\, D^+ \subseteq A_u^+(X))
\geq 1-\alpha\},
\end{equation*}
where the sets $(D^+,D^-)$ are open. The sets $M_{u,\alpha}^+$ and $M_{u,\alpha}^-$ are denoted as the pair of level $u$ avoiding sets. The notion of level avoiding sets can naturally be extended to multiple levels $u_1< u_2 < \cdots < u_k$, which is needed when studying contour maps. In this case, the multilevel contour avoiding set is denoted $C_{\mv{u},\alpha}(X)$ \citep[For a formal definition, see][]{bolin15contours}.

\subsubsection{Excursion functions}
\cite{bolin12} introduced excursion functions as a tool for visualizing excursion sets simultaneously for all values of $\alpha$. For a level $u$, the positive and negative excursion functions are defined as $F_u^+(\mv{s}) = 1 - \inf\{\alpha ; \mv{s}\in \exset{u,\alpha}{+} \}$ and $F_u^-(\mv{s}) = 1 - \inf\{\alpha ; \mv{s}\in \exset{u,\alpha}{-} \}$, respectively. Similarly, the contour avoidance function, and the contour function are defined as $F_u(\mv{s}) = 1 -\inf\{\alpha ; \mv{s}\in \exset{u,\alpha}{} \}$ and
$F_u^c(\mv{s}) = \sup\{\alpha; \mv{s}\in \exset{u,\alpha}{c}\}$, respectively. Finally, for levels $u_1< u_2 < \cdots < u_k$, one can define a contour map function as $F(\mv{s}) = \sup\{1-\alpha; \mv{s}\in C_{u,\alpha}\}$.

These functions take values between zero and one and each set $\exset{u,\alpha}{\bullet}$ can be retrieved as the $1-\alpha$ excursion set of the function $F_u^{\bullet}(\mv{s})$. An example of an excursion set and the corresponding excursion function is shown in Figure~\ref{fig:theory1}.

\subsubsection{Contour maps and their quality measures}
For a function $f(\mv{s})$, a contour map $C_f$ with contour levels $u_1 < u_2 < \ldots < u_K$ is defined as the collection of contour curves $A_{u_1}^c(f), \ldots, A_{u_K}^c(f)$ and associated level sets $\mapset_k = \{\mv{s} : u_{k}< f(\mv{s}) < u_{k+1}\}$, for $0\leq k\leq K$, where one defines $u_0 = -\infty$ and $u_{K+1}=\infty$. In practice, a process $X(\mv{s})$ is often visualized using a contour map of the posterior mean $\E(X(\mv{s})|\mv{Y})$. The contour maps is visualized either by just drawing the contour curves labeled by their values, or by also visualising each level set in a specific color. The color for a set $\mapset_k$ is typically chosen as the color that corresponds to the level $u_k^e = (u_k+u_{k+1})/2$ in a given color map. An example of this is shown in Figure~\ref{fig:theory1}.

In order to choose an appropriate number of contours, one must be able to quantify the uncertainty of contour maps. The uncertainty can be represented using a contour map quality measure $P$, which is a function that takes values in $[0, 1]$. Here, $P$ should be chosen in such a way that $P \approx 1$ indicates that the contour map, in some sense, is appropriate as a description of the distribution of the random field, whereas $P \approx 0$ should indicate that the contour map is inappropriate.

An example of a contour map quality measure is the normalized integral of the contour map function
\begin{equation}
P_0(X, C_f) = \frac1{|\Omega|}\int_{\Omega}F(\mv{s})d\mv{s}.
\end{equation}
The most useful quality measure is denoted $P_2$ and is defined as the simultaneous probability for the level crossings of $(u_1^e, \ldots , u_K^e)$ all falling within their respective level sets $(\mapset_1, \ldots, G_K)$  \cite[for details, see][]{bolin15contours}.

An intuitively interpretable approach for choosing the number of contours in a contour map is to find the largest K such that $P_2$ is above some threshold. For a joint credibility of $90\%$, say, choose the largest number of contours such that $P_2 \geq 0.9$. How this can be done using \pkg{excursions} is illustrated in Section~\ref{sec:precip}.

\subsubsection{Simultaneous confidence bands}
Especially for time series applications, the uncertainty in the latent process is often visualized using pointwise confidence bands. A pointwise confidence interval for $X$ at some location $\mv{s}$ is given by $[q_{\alpha/2}(\mv{s}),q_{1-\alpha/2}(\mv{s})]$, where $q_{\alpha}(\mv{s})$ denotes the $\alpha$-quantile in the marginal distribution of $X(\mv{s})$.

A problem with using pointwise confidence bands is that there is not joint interpretation, and one is therefore often of interested in computing simultaneous confidence bands. For a process $X(\mv{s}), \mv{s}\in \Omega$, we define a simultaneous confidence band as the region $\{(\mv{s},y): \mv{s}\in \Omega, q_{\rho}(\mv{s}) \leq y \leq q_{1-\rho}(\mv{s})\}$. Here $\rho$ is chosen such that $\Prob(q_{\rho}(\mv{s})<X(\mv{s}) <q_{1-\rho}(\mv{s}), \mv{s}\in \Omega)=1-\alpha$. Thus $\alpha$ controls the probability that the process is inside the confidence band at all locations in $\Omega$. An example of pointwise and simultaneous confidence bands is given in Figure~\ref{fig:tokyo1}.

\subsection{Computational methods}
If the latent process $X(\mv{s})$ is defined on a continuous domain $\Omega$, one has to use a discretization, $\mv{x}$, of the process in the statistical analysis. The vector $\mv{x}$ may be the process evaluated at some locations of interest or more generally the weights in a basis expansion $X(\mv{s}) = \sum_i \varphi_i(\mv{s}) x_i$. Computations in the package are in a first stage performed using the distribution of $\mv{x}$. If $\Omega$ is continuous, computations in a second stage interpolates the results for $\mv{x}$ to $\Omega$. In this section, we briefly outline the computational methods used in these two stages. As most quantities of interest can be obtained using some excursion function, we focus on how these are computed. As a representative example, we outline how $F_u^+(\mv{s})$ is computed in the following sections.

As before, let $\mv{Y}$ and $\mv{\theta}$ be a vectors respectively containing observations and model parameters. Computing an excursion function $\mv{F}_u^+ = \{F_u^+(x_1), \ldots, F_u^+(x_n)\}$ requires computing integrals of the posterior distribution for $\mv{x}$. To save computation time, it is assumed that $\exset{u,\alpha_1}{+} \subset \exset{u,\alpha_2}{+}$ if $\alpha_1 > \alpha_2$. This means that $\mv{F}_u$ can be obtained by first reordering the nodes and then computing a sequential integral. The reordering is in this case obtained by sorting the marginal probabilities $\Prob(x_i > u)$ \cite[for other options, see][]{bolin12}.
After reordering, the $i$:th element of $\mv{F}_u^+$ is obtained as the integral
\begin{equation}\label{eq:Fi}
\int_u^{\infty}\pi(\mv{x}_{1:i}|\mv{Y}) d\mv{x}_{1:i}.
\end{equation}
Using sequential importance sampling as described below, the whole sequence of integrals can be obtained with the same cost as computing only one integral with $i=n$, making the computation of $\mv{F}_u^+$ feasible also for large problems.

\subsubsection{Gaussian integrals}
The basis for the computational methods in the package is the ability to compute the required integral when the posterior distribution is Gaussian. In this case, one should compute an integral
\begin{equation}\label{eq:markovint}
\frac{|\mv{Q}|^{1/2}}{(2\pi)^{d/2}}\int_{\mv{u}-\mv{\mu}\leq \mv{x}}\exp\left(-\frac{1}{2}\mv{x}^{\trsp}\mv{Q} \mv{x}\right) \md \mv{x}.
\end{equation}
Here $\mv{\mu}$ and $\mv{Q}$ are the posterior mean and posterior precision matrix respectively. To take advantage of the possible sparsity of $\mv{Q}$ if a Markovian model is used, the integral is rewritten as
\begin{equation}\label{eq:seqint}
\int_{a_d}^{\infty}\pi(x_d)
\int_{a_{d-1}}^{\infty}\pi(x_{d-1}|x_d)
\cdots
\int_{a_2}^{\infty}\pi(x_2|\mv{x}_{3:d})
\int_{a_1}^{\infty} \pi(x_1|\mv{x}_{2:d})
\md \mv{x}
\end{equation}
where, if the problem has a Markov structure, $x_i|\mv{x}_{i+1:d}$ only depends on a few of the elements in $x_{i+1:d}$ given by the Markov structure. The integral is calculated using a sequential importance sampler by starting with the integral of the last component $\pi(x_d)$ and then moving backward in the indices, see \cite{bolin12} for further details.

\subsubsection{Handling non-Gaussian data}
Using the sequential importance sampler above, $\mv{F}_u^+$ can be computed for Gaussian models with known parameters. For more complicated models, the latent Gaussian structure has to be handled, and this can be done in different ways depending on the accuracy that is needed. \pkg{excursions} currently supports the following five methods described in detail in \cite{bolin12}: Empirical Bayes (\code{EB}), Quantile Correction (\code{QC}), Numerical Integration (\code{NI}), Numerical integration with Quantile Corrections (\code{NIQC}), and improved Numerical integration with Quantile Corrections (\code{iNIQC}).

The \code{EB} method is the simplest method and is based on using a Gaussian approximation of the posterior, $\pi(\mv{x}|\mv{Y}) \approx \pi_G(\mv{x}|\mv{Y},\widehat{\mv{\theta}})$. The \code{QC} method uses the same Gaussian approximation but modifies the limits in the integral to improve the accuracy. The three other methods are intended for Bayesian models, where the posterior is obtained by integrating over the parameters,
\begin{equation*}
\pi(\mv{x}\mid\mv{Y}) = \int \pi(\mv{x}\mid\mv{Y},\mv{\theta})\pi(\mv{\theta}\mid\mv{Y})d\mv{\theta}.
\end{equation*}
The \code{NI} method approximates the integration with respect to the parameters as in the INLA method, using a sum of representative parameter configurations, and the \code{NIQC} and \code{iNIQC} methods combines this with the \code{QC} method to improve the accuracy further.

In general, \code{EB} and \code{QC} are suitable for frequentist models and for Bayesian models where the posterior distribution conditionally on the parameters is approximately Gaussian. The methods are equivalent if the posterior is Gaussian and in other cases \code{QC} is more accurate than \code{EB}. For Bayesian models, the \code{NI} method is, in general, more accurate than the \code{QC} method, and for non-Gaussian likelihoods, the \code{NIQC} and \code{iNIQC} methods can be used for improved results. In general the accuracy and computational cost of the methods are as follows
\begin{align*}
\mbox{accuracy: } & \mbox{\code{EB} $<$ \code{QC} $<$ \code{NI} $<$ \code{NIQC} $<$ \code{iNIQC}} \\
\mbox{comp.~cost: } & \mbox{\code{EB} $\approx$ \code{QC} $<$ \code{NI} $\approx$ \code{NIQC} $<$ \code{iNIQC}}.
\end{align*}
If the main purpose of the analysis is to construct excursion or contour sets for low values of $\alpha$, we recommend using the \code{QC} method for problems with Gaussian likelihoods and the \code{NIQC} method for problems with non-Gaussian likelihoods. The increase in accuracy of the \code{iNIQC} method is often small compared to the added computational cost.

\subsubsection{Continuous domain interpolations}
For a continuous spatial domain, the excursion function $F_u^+(\mv{s})$  can be approximated using $\mv{F}_u^+$ computed at discrete point locations. The main idea is to interpolate $\mv{F}_u^+$ assuming monotonicity of the random field between the discrete computation locations. Specifically, assume that the values of $\mv{F}_u^+$ correspond to the values at the vertices of some triangulated mesh such as the one shown in the left panel of Figure~\ref{fig:theory1}. If the excursion set $\exset{u,\alpha}{+}(X)$ should be computed for some specific value of $\alpha$, one has to find the $1-\alpha$ contour for $F_u^+(\mv{s})$. For interpolation to a location $\mv{s}$ within a specific triangle $\mathcal{T}$ with corners in $\mv{s}_1, \mv{s}_2,$ and $\mv{s}_3$, \pkg{excursions} by default uses log-linear interpolation, $F_u(\mv{s}) = \exp\{\sum_{k=1}^3 w_k\log[F_u(\mv{s}_k)]\}$. Here $\{(w_1,w_2,w_3);\, w_1,w_2,w_3\geq 0,\, \sum_{k=1}^3 w_i=1\}$ are the barycentric coordinates of $\mv{s}$ within the triangle.

Further technical details of the continuous domain construction are given in   \cite{bolin15contours}. Studies of the resulting continuous domain excursion sets in \cite{bolin15contours} indicate that the log-linear interpolation method results in sets with coverage probability on target or slightly above target for large target probabilities. An example of a continuous domain excursion set for a triangulated mesh is shown in the middle panel of Figure~\ref{fig:theory1}. In the right panel of the figure, the interpolated function $F_u^+(\mv{s})$ is shown. The code that generates the figure is explained in the next section.

\section{Implementation}\label{sec:software}
The functions in \pkg{excursions} can be divided into four main categories depending on what they compute: (1) Excursion sets and credible regions for contour curves, (2) Quality measures for contour maps, (3) Simultaneous confidence bands, and (4) Utility such as Gaussian integrals and continuous domain mappings. The main functions come in at least three different versions taking different input: (1) The parameters of a Gaussian process, (2) results from an analysis using the \proglang{R}-\pkg{INLA} software package, and (3) Monte Carlo simulations of the process. These different categories are described in further detail below.

Much of the computations in the package is done in C functions. These functions use methods from a number of C and Fortran libraries, such as \code{BLAS} \citep{blas90}, \code{LAPACK} \citep{lapack99}, and \code{CHOLMOD} \citep{cholmod08} for efficient matrix manipulations together with function from the GNU Scientific library \citep{GSL06} and several different reordering methods. Notably, the \code{CAMD} library \citep{camd96,camd04} is used for constrained approximate minimum degree orderings.

As an example that will be used to illustrate the methods in later sections, we generate data $Y_i \sim \pN(X(\mv{s}_i),\sigma^2)$ at some locations $\mv{s}_1, \ldots, \mv{s}_{100}$ where $X(\mv{s})$ is a Gaussian random field specified using a stationary SPDE model \citep{lindgren10}.

\begin{Schunk}
\begin{Sinput}
R> x <- seq(from = 0, to = 10, length.out = 20)
R> mesh <- inla.mesh.create(lattice = inla.mesh.lattice(x = x, y = x),
+   extend = FALSE, refine = FALSE)
R> spde <- inla.spde2.matern(mesh, alpha = 2)
R> obs.loc <- 10 * cbind(runif(100), runif(100))
R> Q <- inla.spde2.precision(spde, theta = c(log(sqrt(0.5)), 0))
R> x <- inla.qsample(Q = Q, seed = seed)
\end{Sinput}
\end{Schunk}

Based on the observations, we calculate the posterior distribution of the latent field, which is Gaussian with mean \code{mu.post} and precision matrix \code{Q.post}, these are computed as follows. We refer to \cite{lindgren2015software} for details about the \proglang{R}-\pkg{INLA} related details in the code.

\begin{Schunk}
\begin{Sinput}
R> A <- inla.spde.make.A(mesh = mesh, loc = obs.loc)
R> sigma2.e = 0.01
R> Y <- as.vector(A %*% x + rnorm(100) * sqrt(sigma2.e))
R> Q.post <- (Q + (t(A) %*% A) / sigma2.e)
R> mu.post <- as.vector(solve(Q.post, (t(A) %*% Y) / sigma2.e))
\end{Sinput}
\end{Schunk}

\begin{figure}[t]
\centerline{%
\includegraphics[trim= 75 77 60 70,clip,width=0.3\linewidth]{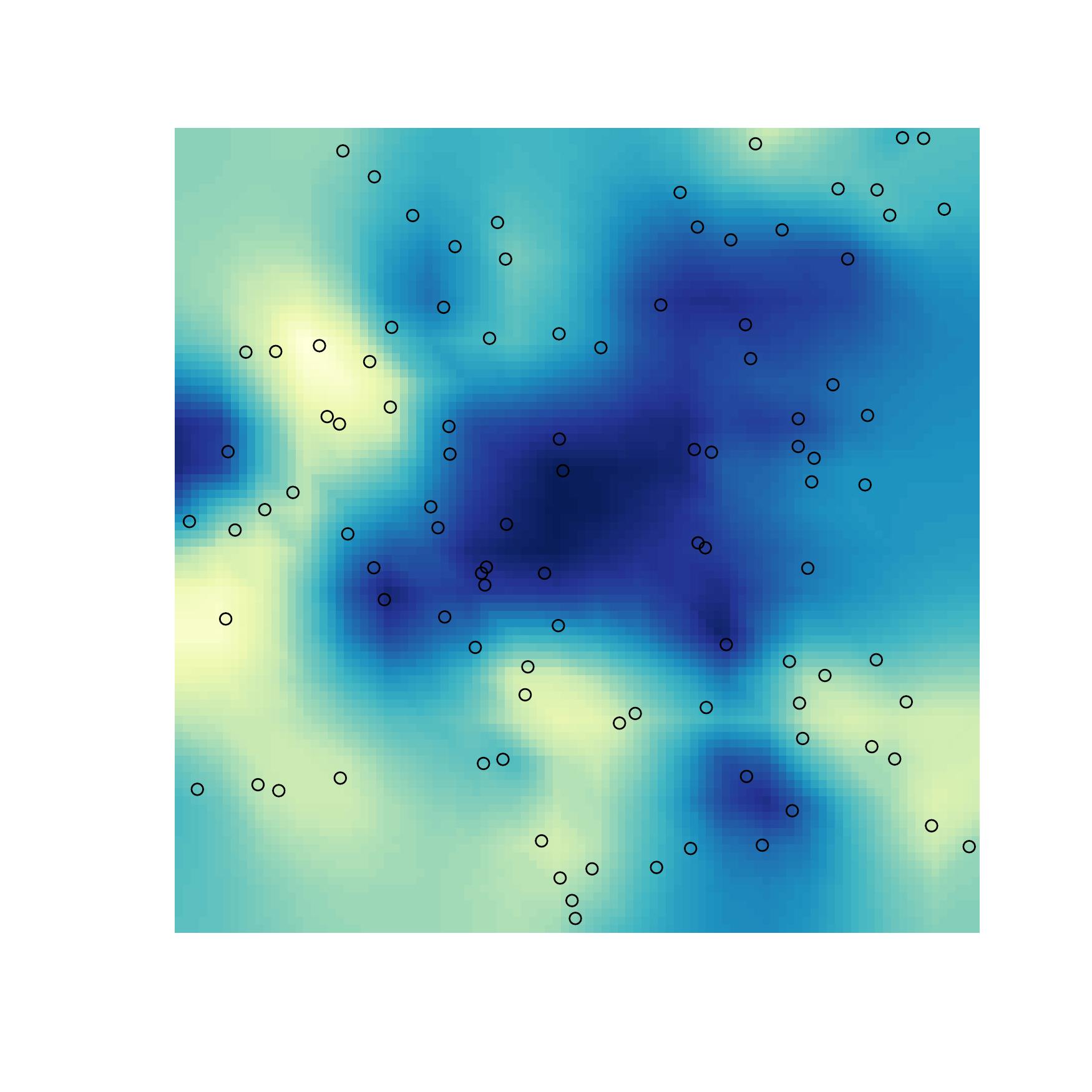}
\hspace*{-0.15cm}\raisebox{-0.95cm}{\includegraphics[trim= 426 0 0 0,clip, width=0.063\linewidth]{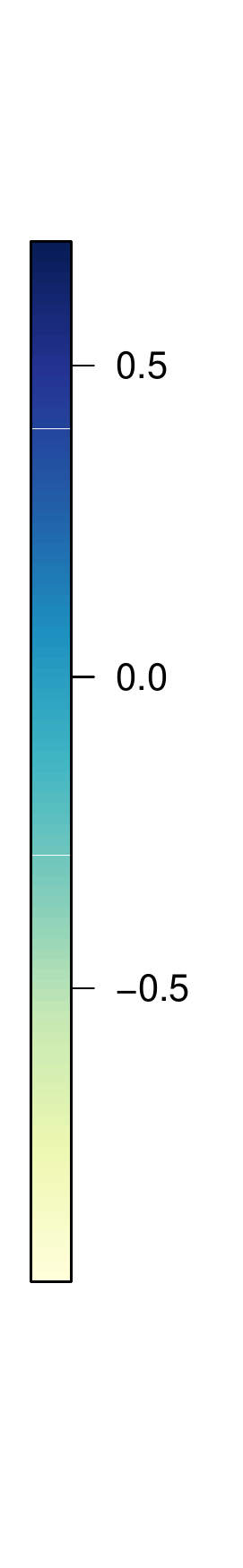}}
\includegraphics[trim= 75 77 60 70,clip,width=0.3\linewidth]{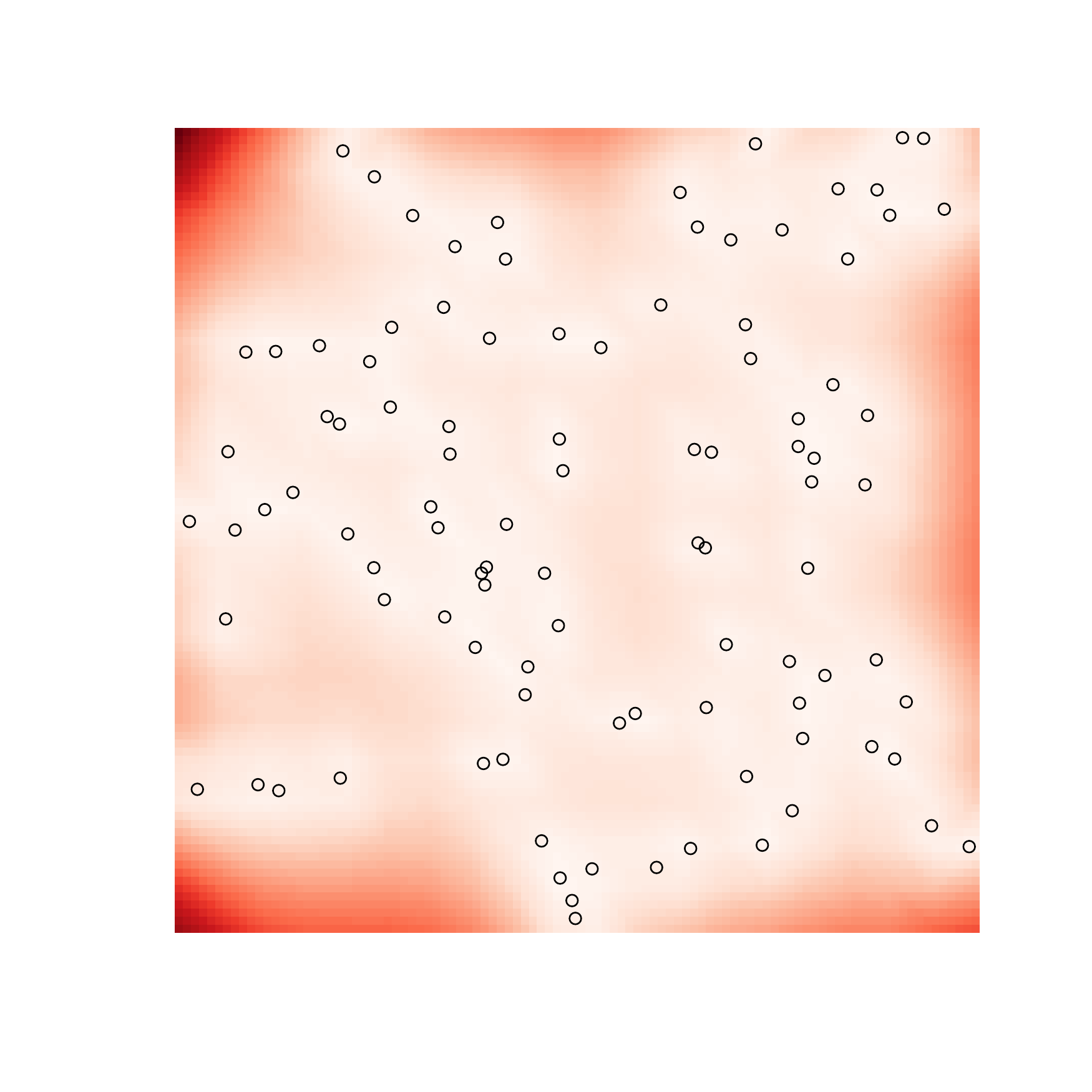}
\hspace*{-0.15cm}\raisebox{-0.95cm}{\includegraphics[trim= 426 0 0 0,clip, width=0.063\linewidth]{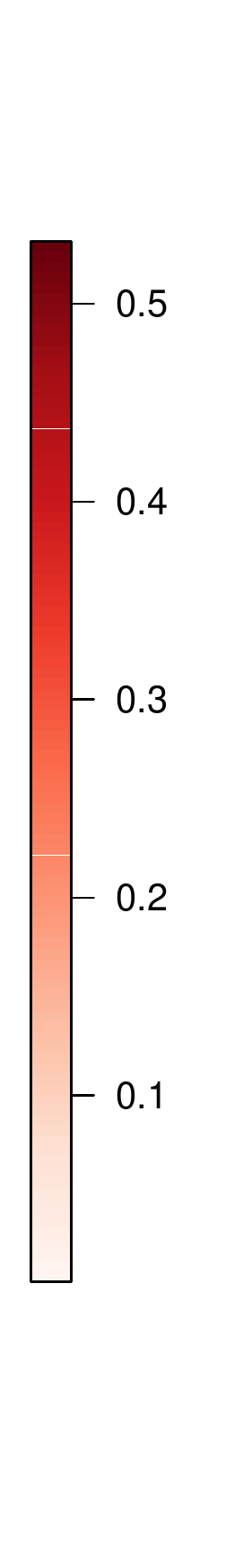}}}
\vspace{-1cm}
\caption{Posterior mean (left) and posterior standard deviations of an example using simulated data. The measurement locations are marked with circles in both panels. }
\label{fig:theory0}
\end{figure}

\begin{figure}[t]
\centerline{\includegraphics[trim= 75 77 60 70,clip,width=0.3\linewidth]{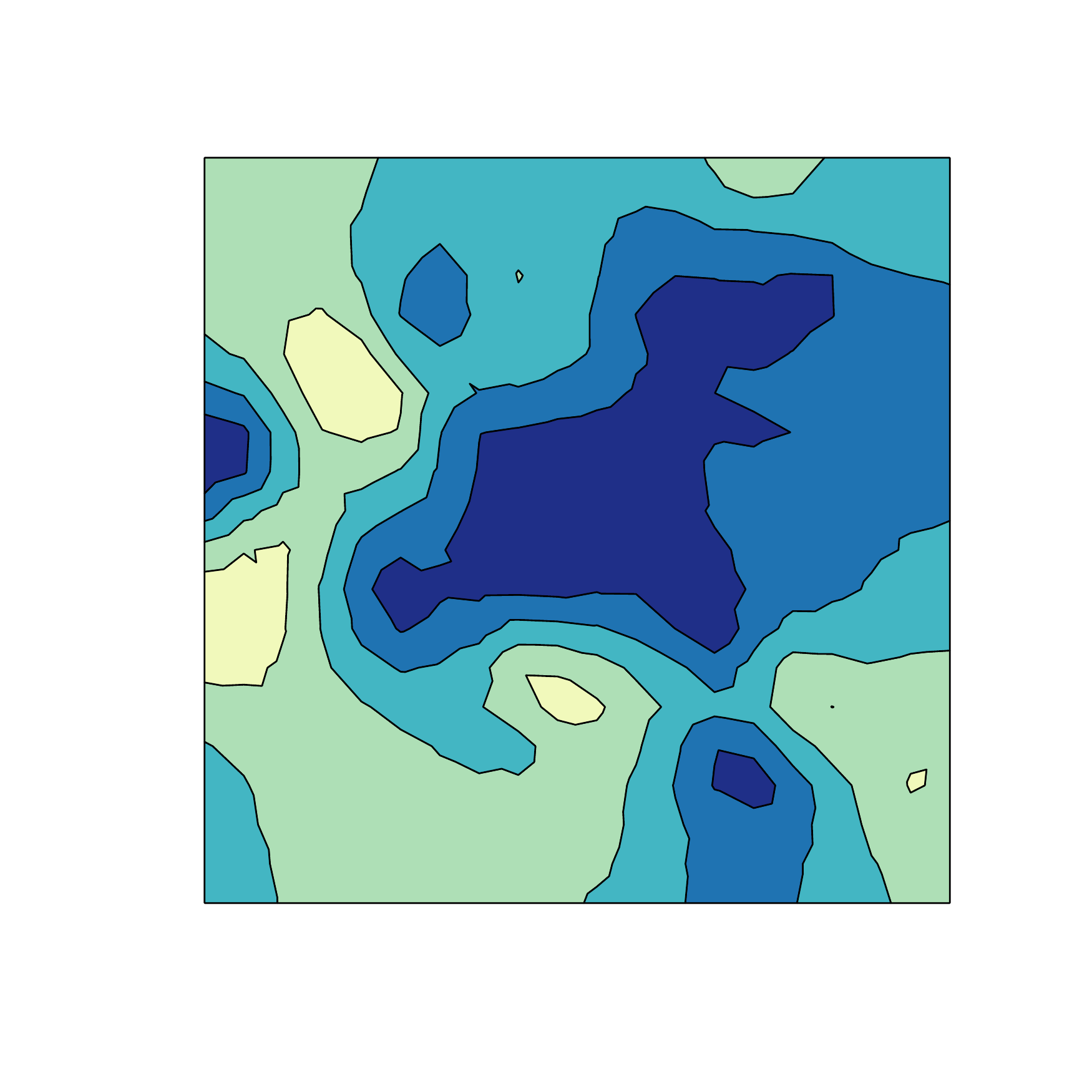}
\hspace*{-0.15cm}\raisebox{-0.95cm}{\includegraphics[trim= 426 0 0 0,clip, width=0.063\linewidth]{theorymapcm-1}}
\includegraphics[trim= 77 77 60 70,clip,width=0.3\linewidth]{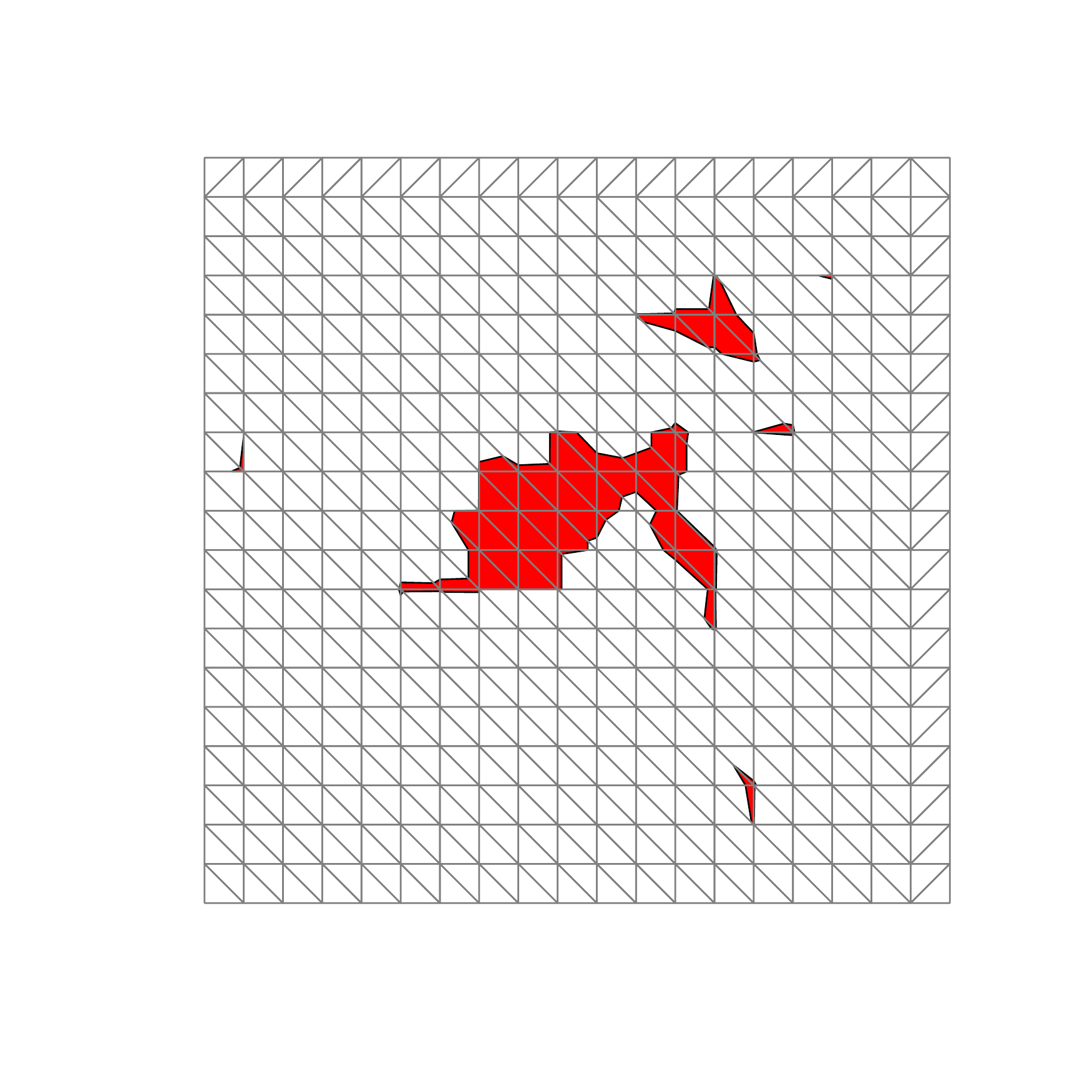}
\includegraphics[trim= 60 60 50 60,clip,width=0.3\linewidth]{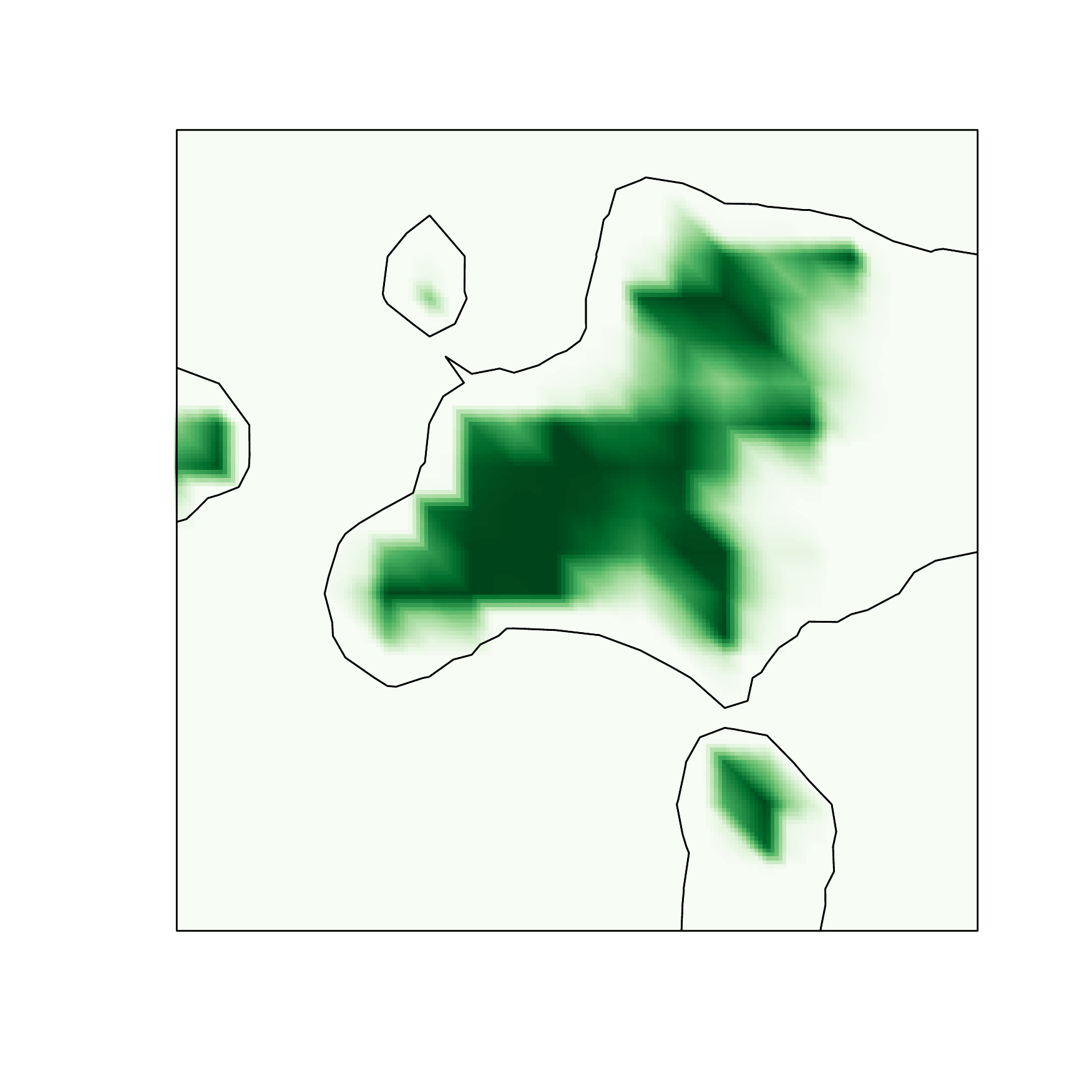}
\raisebox{-0.7cm}{\includegraphics[trim= 434 15 0 0,clip,width=0.0565\linewidth]{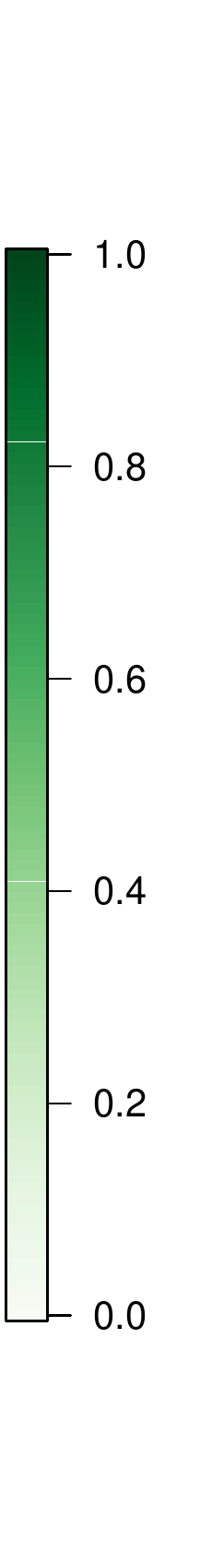}}}
\vspace{-1cm}
\caption{Contour map of the posterior mean shown in Figure~\ref{fig:theory0} (left), a triangulated mesh and interpolated excursion set $\exset{0,0.1}{+}$ (mid), and the interpolated excursion function $F_0^+(s)$ (right). In the right panel,  the zero-level contour is also plotted.}
\label{fig:theory1}
\end{figure}

Figure~\ref{fig:theory0} shows the posterior mean and the posterior standard deviations. A contour map of the posterior mean is shown in Figure~\ref{fig:theory1}.

\subsection{Excursion sets and contour credible regions}
The main function for computing excursion sets and contour credible regions is  \texttt{excursions}. A typical call to the function looks like
\begin{Schunk}
\begin{Sinput}
R> res.exc <-  excursions(mu = mu.post, Q = Q.post, alpha = 0.1, type = '>',
+   u = 0, F.limit = 1)
\end{Sinput}
\end{Schunk}

Here, \code{mu} and \code{Q} are the mean vector and precision matrix for the joint distribution of the Gaussian vector \code{x}. The arguments \code{alpha}, \code{u}, and \code{type} are used to specify what type of excursion set that should be computed: \code{alpha} is the error probability, \code{u} is the excursion or contour level, and \code{type} determines what type of region that is considered: '>' for positive excursion regions, '<' for negative excursion regions, '!=' for contour avoiding regions, and '=' for contour credibility regions. Thus, the call above computes the excursion set $\exset{0,0.1}{+}$.

The argument \code{F.limit} is used to specify when to stop the computation of the excursion function. In this case with \code{F.limit=1}, all values of $\mv{F}_u^+$ are computed, but the computation time can be reduced by decreasing the value of \code{F.limit}.

The function has the EB method as default strategy for handling the possible latent Gaussian structure. In the simulated example, the likelihood is Gaussian and the parameters are assumed to be known, so the EB method is exact. The QC method can be used instead by specifying \code{method='QC'}. In this case, the argument \code{rho} should be used to also provide a vector with point-wise marginal probabilities: $P(x_i>u)$ for positive excursions and contour regions, and $P(x_i<u)$ for negative excursions. In the situation when $\pi(\mv{x}|\mv{Y},\mv{\theta})$ is Gaussian but $\pi(\mv{x}|\mv{Y})$ is not, the marginal probabilities should be calculated under the distribution $\pi(\mv{x}|\mv{Y})$ and \code{mu} and \code{Q} should be chosen as the mean and precision for the distribution $\pi(\mv{x}|\mv{Y},\hat{\mv{\theta}})$ where $\hat{\mv{\theta}}$ is the MAP or ML estimate of the parameters.

\subsubsection{The INLA interface}
The function \code{excursions.inla} is used to compute excursion sets and credible regions for contour curves for latent Gaussian models that have been estimated using \proglang{R}-\pkg{INLA}. It takes the same arguments \code{excursions}, except that \code{mu} and \code{Q} are replaced with arguments related to \proglang{R}-\pkg{INLA}. A basic call to the function \code{excursions.inla} looks like
\begin{Schunk}
\begin{Sinput}
R> excursions.inla(result.inla, name, alpha, u, type)
\end{Sinput}
\end{Schunk}
Here \code{result.inla} is the output from a call to the \code{inla} function, and \code{name} is the name of the component in the output that the computations should be done for. For more complicated models, one typically specifies the model in \proglang{R}-\pkg{INLA} using an \code{inla.stack} object. If this is done, the call to \code{excursions.inla} will instead look like
\begin{Schunk}
\begin{Sinput}
R> excursions.inla(result.inla, stack, tag, alpha, u, type)
\end{Sinput}
\end{Schunk}
Here \code{stack} is now the stack object and \code{tag} is the name of the component in the stack for which the computations should be done. The typical usage of the \code{tag} argument is to have one part of the stack that contains the locations where measurements are taken, and another that contains the locations where the output should be computed.

\code{excursions.inla} has support for all strategies described in Section~\ref{sec:methods} for handling latent Gaussian structures: The argument \code{method} can be one of \code{'EB'}, \code{'QC'}, \code{'NI'}, \code{'NIQC'}, and \code{'iNIQC'}.

\subsubsection{Analysis of Monte Carlo samples}
The function \texttt{excursions.mc} can be used to post-process Monte Carlo model simulations in order to compute excursion sets and credible regions. For this function, the model is not specified explicitly. Instead a $d \times N$ matrix \code{X} containing $N$ Monte Carlo simulations of the $d$ dimensional process of interest is provided. A basic call to the function looks like
\begin{Schunk}
\begin{Sinput}
R> excursions.mc(X, u, type)
\end{Sinput}
\end{Schunk}
where \code{u} again determines the level of interest and \code{type} defines the type of set that should be computed. It is important to note that this function does all computations purely based on the Monte Carlo samples that are provided, and it does not use any of the computational methods based on sequential importance sampling for Gaussian integrals that the is the basis for the previous methods. This means that this function in one sense is more general as \code{X} can be samples from any model, not necessarily a latent Gaussian model. The price that has to be payed for this generality is that the only way of increasing the accuracy of the results is to increase the number of Monte Carlo samples that are provided to the function.

\subsection{Analysis of contour maps}
The main function for analysis of contour maps is \code{contourmap}. A basic call to the function looks like
\begin{Schunk}
\begin{Sinput}
R> res.con <- contourmap(mu = mu.post, Q = Q.post, n.levels = 4,
+   alpha = 0.1, compute = list(F = TRUE, measures = c("P0")))
\end{Sinput}
\end{Schunk}

Here, \code{mu} is again the mean value and \code{Q} is the precision matrix of the model. The parameter \code{n.levels} sets the number of contours that should be used in the contour map, and these are spaced equidistant in the range of \code{mu} by default. Other types of contour maps can be obtained using the \code{type} argument. For manual specification of the levels, the \code{levels} argument can be used instead. By default, the function computes the specified contour map but no quality measures and it does not compute the contour map function. If quality measures should be computed, this is specified using the \code{compute} argument. This argument is also used to decide whether the contour map function $F$ should be computed.

As for \code{excursions}, this function comes in two other versions depending on the form of the input:  \code{contourmap.inla} for model specification using an \proglang{R}-\pkg{INLA} object, or \code{contourmap.mc} for model specification using Monte Carlo simulations of the model. The model specification using these functions is identical to that in the corresponding \code{excursions} functions.

\subsection{Continuous domain interpretations}

A common scenario is that the input used in \code{contourmap} or \code{excursions} represents the value of the model at some discrete locations in a continuous domain of interest. In this case, the function \code{continuous} can be used to interpolate the discretely computed values by assuming monotonicity of the random field in between the discrete computation locations, as discussed in Section~\ref{sec:methods}. A typical calls to the function looks like
\begin{Schunk}
\begin{Sinput}
R> sets.exc <- continuous(ex = res.exc, geometry = mesh, alpha = 0.1)
\end{Sinput}
\end{Schunk}
Here \code{ex} is the result of the call to \code{contourmap} or \code{excursions} and \code{alpha} is the error probability of interest for the excursion set or credible region computation. The argument \code{geometry} specifies the geometric configuration of the values in input \code{ex}, either as a general triangulation geometry or as a lattice. A lattice can be specified as an object of the form \code{list(x, y)} where \code{x} and \code{y} are vectors, or as \code{list(loc, dims)} where \code{loc} is a two-column matrix of coordinates, and \code{dims} is the lattice size vector. If \proglang{R}-\pkg{INLA} is used, the lattice can also be specified as an \code{inla.mesh.lattice} object. In all cases, the input is treated topologically as a lattice with lattice boxes that are assumed convex. A triangulation geometry is specified as an \code{inla.mesh} object. Finally, an argument \code{output} can be used to specify what type of object should be generated. The options are currently \code{sp} which gives a \code{SpatialPolygons} \citep{bivand2013applied} object, and \code{inla} which gives an \code{inla.mesh.segment} object.

\subsection{Simultaneous confidence bands}
The function \code{simconf} can be used for calculating simultaneous confidence bands for a Gaussian process $X(s)$. A basic call to the function looks like
\begin{Schunk}
\begin{Sinput}
R> simconf(alpha, mu, Q)
\end{Sinput}
\end{Schunk}
where \code{alpha} is the coverage probability, \code{mu} is the mean value vector for the process, and \code{Q} is the precision matrix for the process. The function has a few optional arguments similar to those of \code{excursions}, all listed in the documentation of the function. The function returns upper and lower limits for both pointwise and simultaneous confidence bands.

As for \code{excursions} and \code{contourmap}, there is also a version of \code{simconf} that can be used to analyze \proglang{R}-\pkg{INLA} models (\code{simconf.inla}) and a version that can analyze Monte Carlo samples (\code{simconf.mc}). Furthermore, there is a version \code{simconf.mixture} which is used to compute simultaneous confidence regions for Gaussian mixture models with a joint distribution on the form
$$
\pi(x) = \sum_{k=1}^K w_k \pN(\mu_k, Q_k^{-1}).
$$
This particular function was used to analyze the models in \cite{bolin2014statistical} and \cite{guttorp2014assessing}, but is also used internally by \code{simconf.inla}.

\subsection{Gaussian integrals}
Among the utility functions in the package, the function \code{gaussint} can be especially useful also in a larger context. It contains the implementation of the sequential importance sampling method for computing Gaussian integrals, described in Section~\ref{sec:methods}. This function has two features that separates it from many other functions for computing Gaussian integrals: Firstly it is based on the precision matrix of the Gaussian distribution, and sparsity of this matrix can be utilized to decrease computation time. Secondly, the integration can be stopped as soon as the value of the integral in the sequential integration goes below some given value $1-\alpha$. If one only is interested in the exact value of the integral given that it is larger than some value $1-\alpha$, this option can save a lot of computation time.

A basic call to the function looks like
\begin{Schunk}
\begin{Sinput}
R> gaussint(mu, Q, a, b)
\end{Sinput}
\end{Schunk}
where \code{mu} is the mean value vector, \code{Q} is the precision matrix, \code{a} is a vector of the lower limits in the integral, and \code{b} contains the upper integration limits. If the Cholesky factor of \code{Q} is known beforehand, this can be supplied to the function using the \code{Q.chol} argument. An argument \code{alpha} is used to set the computational $1-\alpha$ limit for the integral. The function returns an estimate of the integral as well as an error estimate. If the error estimate is too high, the precision can be increased by increasing the \code{n.iter} argument of the function.

\subsection{Plotting}
The \pkg{excursions} package contains various functions that are useful for visualization. The function \code{tricontourmap} can be used for visualization of contour maps computed on triangulated meshes. The following code plots the posterior mean using the contour map we previously computed.
\begin{Schunk}
\begin{Sinput}
R> set.sc <- tricontourmap(mesh, z = mu.post, levels = res.con$u)
R> cmap <- colorRampPalette(brewer.pal(9, "YlGnBu"))(100)
R> cols <- contourmap.colors(res.con, col = cmap)
R> plot(set.sc$map, col = cols)
\end{Sinput}
\end{Schunk}
Here \code{contourmap.colors} is used to find appropriate colors for each set in the contour map, based on the color map \code{cmap} that was defined using the \pkg{RColorBrewer} \citep{RColorBrewer} package.  The results of the following commands are shown in Figure~\ref{fig:theory1}. The estimated excursion set $\exset{u,\alpha}{+}(X)$, can be visualized as
\begin{Schunk}
\begin{Sinput}
R> plot(sets.exc$M["1"], col = "red", xlim = range(mesh$loc[,1]),
+   ylim = range(mesh$loc[,2]))
R> plot(mesh, vertex.color = rgb(0.5, 0.5, 0.5), draw.segments = FALSE,
+   edge.color = rgb(0.5, 0.5, 0.5), add = TRUE)
\end{Sinput}
\end{Schunk}
The second \code{plot} command adds the mesh to the plot so that we can see how the set is interpolated by the \code{continuous} function. Finally, the interpolated excursion function $F_u^+(\mv{s})$, can be plotted easily using the \code{inla.mesh.projector} function from the \proglang{R}-\pkg{INLA} package.
\begin{Schunk}
\begin{Sinput}
R> cmap.F <- colorRampPalette(brewer.pal(9, "Greens"))(100)
R> proj <- inla.mesh.projector(sets.exc$F.geometry, dims = c(200, 200))
R> image(proj$x, proj$y, inla.mesh.project(proj, field = sets.exc$F),
+   col = cmap.F, axes = FALSE, xlab = "", ylab = "", asp = 1)
R> con <- tricontourmap(mesh, z = mu.post, levels = 0)
R> plot(con$map, add = TRUE)
\end{Sinput}
\end{Schunk}
The final two lines computes the level zero contour curve and plots it in the same figure as the interpolated excursion function. The colorbars in the figures are plotted using the \pkg{fields} \citep{nychka2015fields} package.

\section{Two applications}\label{sec:tutorial}
In this section, two examples are used to illustrate how \pkg{excursions} can be used.

\subsection{Time series data: Tokyo rainfall}
\label{sec:tokyo}

To illustrate the methods in the package, we use the much analyzed binomial time series from \cite{kitagawa1987non}. Each day during the years 1983 and 1984, it was recorded whether there was more than 1 mm rainfall in Tokyo. Of interest is to study the underlying probability of rainfall as a function of day of the year. The data is modelled as $y_i \sim Bin(n_i,p_i)$ for calendar day $i = 1,...,366$. Here $n_i = 2$ for all days except for February 29 ($i=60$) which only occurred during the leap year of 1984. The probability $p_i$ is modeled as a logit-transformed Gaussian process.

The model and the following \proglang{R}-\pkg{INLA} implementation of the model is described further in \cite{lindgren2015software}.

\begin{Schunk}
\begin{Sinput}
R> data("Tokyo")
R> mesh <- inla.mesh.1d(seq(1, 367, length = 25), interval = c(1, 367),
+   degree = 2, boundary = "cyclic")
R> kappa <- 1e-3
R> tau <- 1 / (4 * kappa^3)^0.5
R> spde <- inla.spde2.matern(mesh, constr = FALSE, theta.prior.prec = 1e-4,
+   B.tau = cbind(log(tau), 1), B.kappa = cbind(log(kappa), 0))
R> A <- inla.spde.make.A(mesh, loc = Tokyo$time)
R> time.index <- inla.spde.make.index("time", n.spde = spde$n.spde)
R> stack <- inla.stack(data = list(y = Tokyo$y, link = 1, Ntrials = Tokyo$n),
+   A = list(A), effects = list(time.index), tag = "est")
R> formula <- y ~ -1 + f(time, model = spde)
R> data <- inla.stack.data(stack)
\end{Sinput}
\end{Schunk}

Next, the model is estimated using the \code{inla} function. Since we want to analyze the output using \pkg{excursions}, the additional option \code{control.compute = list(config = TRUE)} must be specified in the \code{inla} function. This makes the function save some extra output needed by \pkg{excursions}.

\begin{Schunk}
\begin{Sinput}
R> result <- inla(formula, family = "binomial", data = data,
+   Ntrials = data$Ntrials, control.predictor = list(
+   A = inla.stack.A(stack), link = data$link, compute = TRUE),
+   control.compute = list(config = TRUE))
\end{Sinput}
\end{Schunk}

We now have estimates of the posterior mean and marginal confidence intervals for the probability of rain for each day. However, if we also want joint confidence bands, we can estimate these using \code{simconf.inla} as
\begin{Schunk}
\begin{Sinput}
R> res <- simconf.inla(result, stack, tag = "est", alpha = 0.05, link = TRUE)
\end{Sinput}
\end{Schunk}

\begin{figure}[t]
\centerline{\includegraphics[width=0.6\linewidth,height=0.5\linewidth]{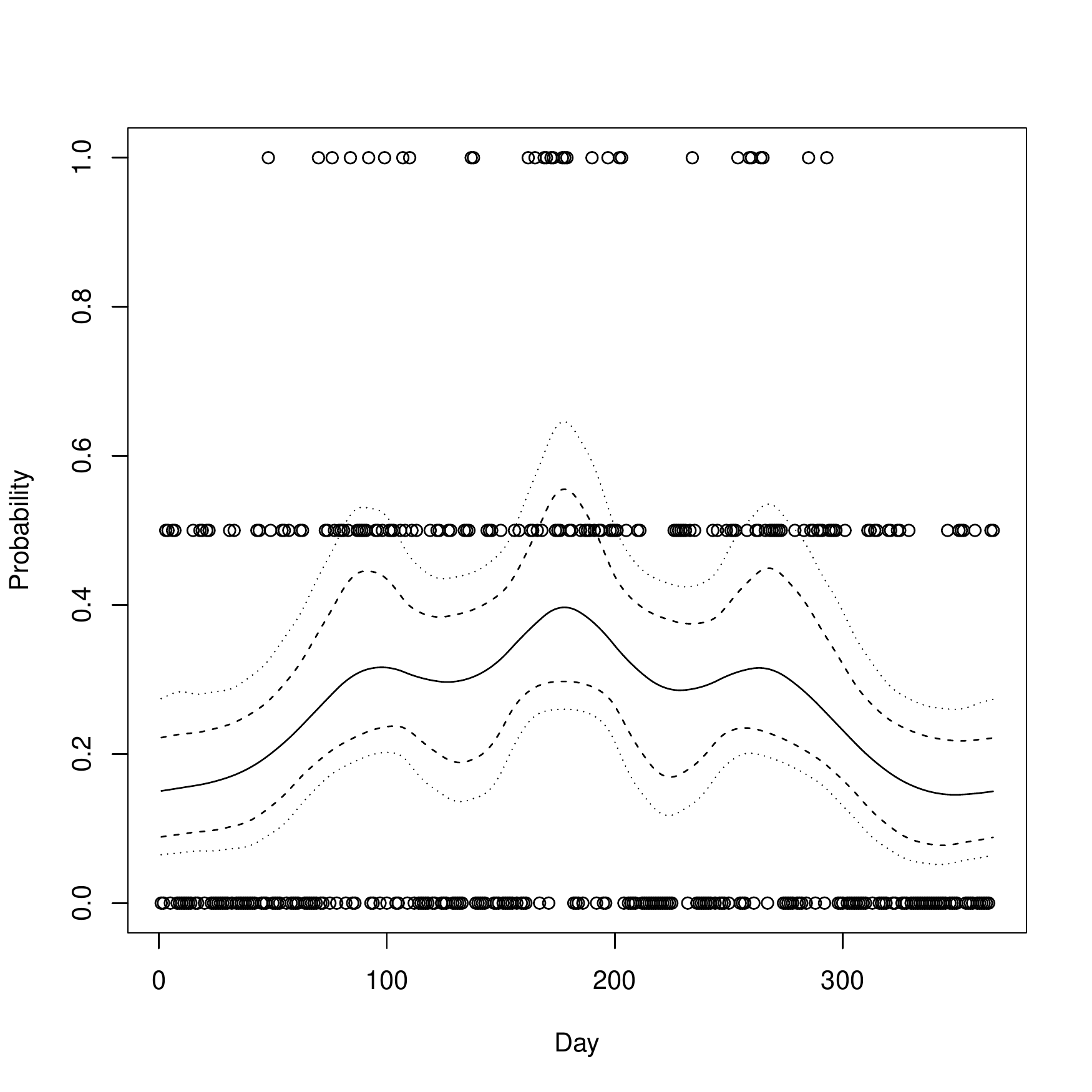}}
\vspace{-0.5cm}
\caption{Empirical and model-based Binomial probability estimates for the Tokyo rainfall data set, with a marginal $95\%$ confidence band (dashed) and a simultaneous confidence band (dotted). The empirical probability estimates are the proportion of observed rainfall days for each day of the year.}
\label{fig:tokyo1}
\end{figure}

Note the argument \code{link=TRUE} which tells the function that the results should be returned in the scale of the data, and not in the scale of the linear predictor. Next, we plot the results, showing the marginal confidence bands with dashed lines and the simultaneous confidence band with dotted lines. The results are shown in Figure~\ref{fig:tokyo1}.

\begin{Schunk}
\begin{Sinput}
R> index <- inla.stack.index(stack, "est")$data
R> plot(Tokyo$time, Tokyo$y / Tokyo$n, xlab = "Day", ylab = "Probability")
R> lines(result$summary.fitted.values$mean[index])
R> matplot(cBind(res$a.marginal, res$b.marginal), type = "l", lty = 2,
+   col = 1, add = TRUE)
R> matplot(cBind(res$a, res$b), type = "l", lty = 3, col = 1, add = TRUE)
\end{Sinput}
\end{Schunk}

\subsection{Spatial data: Parana precipitation}
\label{sec:precip}
We will now use a spatial dataset to illustrate how \pkg{excursions} can be used. In order to keep the parts of the code that are not relevant to \pkg{excursions} brief, we again use data available in \proglang{R}-\pkg{INLA}. The dataset consists of daily rainfall data for each day of the year 2011, at 616 locations in and around the stated of Paran\'a in Brazil. In the following analysis, we analyse the data from the month of January.

The statistical model used for the data is a latent Gaussian model, where the precipitation measurements are assumed to be $\Gamma$-distributed with a spatially varying mean. The mean is modeled as a log-Gaussian Mat\'ern field specified as an SPDE model. Details of the current model model and the following INLA implementation can be found in the SPDE tutorial available on the \proglang{R}-\pkg{INLA} homepage, see \cite{Wallin15} for an analysis of the data using a different non-Gaussian SPDE model.

We start by loading the data and defining the model:

\begin{Schunk}
\begin{Sinput}
R> data("PRprec")
R> data("PRborder")
R> Y <- rowMeans(PRprec[, 3 + (1 : 31)])
R> ind <- !is.na(Y)
R> Y <- Y[ind]
R> coords <- as.matrix(PRprec[ind, 1 : 2])
R> b <- inla.nonconvex.hull(coords, -0.03, -0.05, resolution = c(100, 100))
R> prmesh <- inla.mesh.2d(boundary = b, max.edge = c(.45, 1), cutoff = 0.2)
R> A <- inla.spde.make.A(prmesh, loc = coords)
R> spde <- inla.spde2.matern(prmesh, alpha = 2)
R> mesh.index <- inla.spde.make.index(name = "field", n.spde = spde$n.spde)
\end{Sinput}
\end{Schunk}

The measurement stations are spatially irregular, but we are interested in making predictions to a regular lattice within the state. In order to do continuous domain interpretations, we define the lattice locations as a lattice object using the \code{submesh} function of the \code{excursions} package. The function \code{inout} from the package \pkg{splancs} \citep{spancs_manual} is used to find the locations on the lattice that are within the region of interest.

\begin{Schunk}
\begin{Sinput}
R> nxy <- c(50, 50)
R> projgrid <- inla.mesh.projector(prmesh, xlim = range(PRborder[, 1]),
+   ylim = range(PRborder[, 2]), dims = nxy)
R> xy.in <- inout(projgrid$lattice$loc, cbind(PRborder[, 1], PRborder[, 2]))
R> submesh = submesh.grid(matrix(xy.in, nxy[1], nxy[2]),
+   list(loc=projgrid$lattice$loc, dims = nxy))
\end{Sinput}
\end{Schunk}

We now define the \code{stack} objects and estimate the model using \code{inla}. Again note that we have to set the \code{control.compute} argument of the results is to be used by \pkg{excursions}.
\begin{Schunk}
\begin{Sinput}
R> A.prd <- inla.spde.make.A(prmesh, loc = submesh$loc)
R> stk.prd <- inla.stack(data=list(y = NA), A = list(A.prd, 1),
+   effects=list(c(mesh.index, list(Intercept = 1)),
+   list(lat = submesh$loc[,2], lon = submesh$loc[,1])), tag = "prd")
R> stk.dat <- inla.stack(data = list(y = Y), A=list(A, 1),
+   effects = list(c(mesh.index, list(Intercept = 1)),
+   list(lat = coords[,2], lon = coords[,1])), tag = "est")
R> stk <- inla.stack(stk.dat, stk.prd)
R> r <- inla(y ~ -1 + Intercept + f(field, model = spde), family = "Gamma",
+   data = inla.stack.data(stk), control.compute = list(config = TRUE),
+   control.predictor = list(A = inla.stack.A(stk), compute = TRUE, link=1))
\end{Sinput}
\end{Schunk}

\begin{figure}[t]
\centerline{\includegraphics[trim= 60 60 40 60,clip,width=0.44\linewidth]{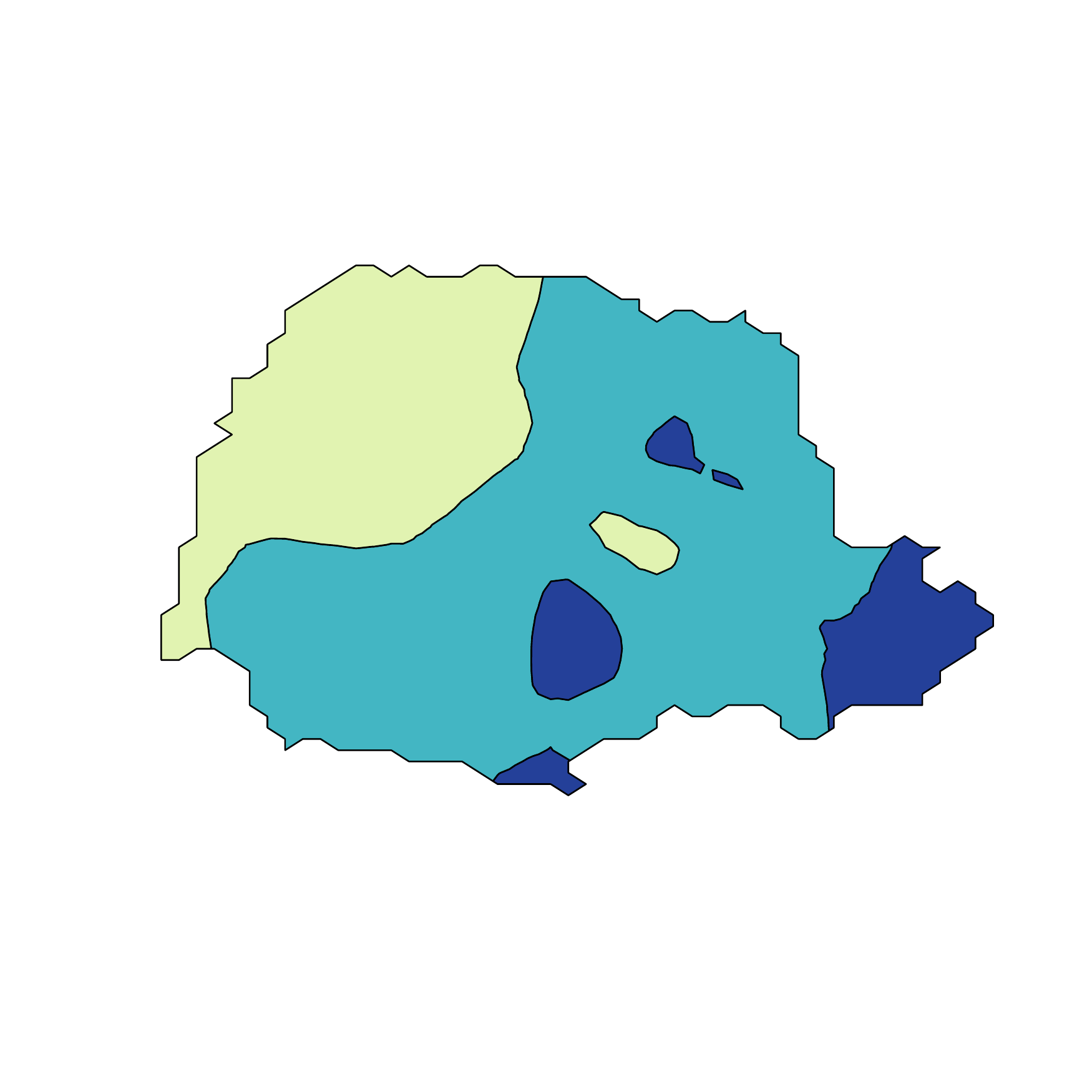}
\raisebox{1cm}{\includegraphics[trim= 434 15 0 0,clip,width=0.05\linewidth]{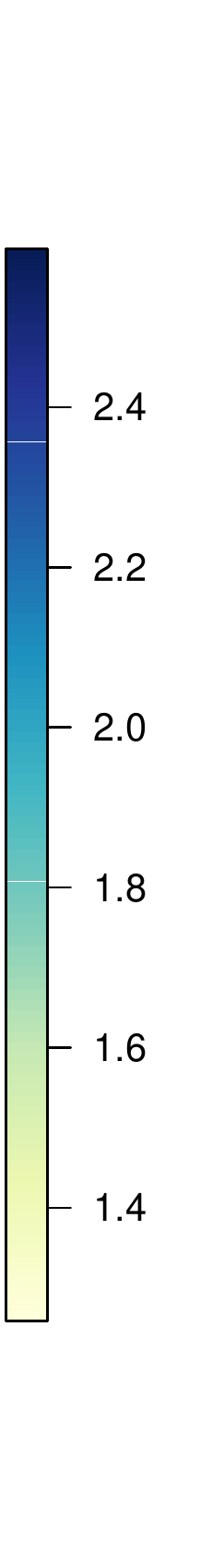}}
\includegraphics[trim= 50 50 30 50,width=0.44\linewidth]{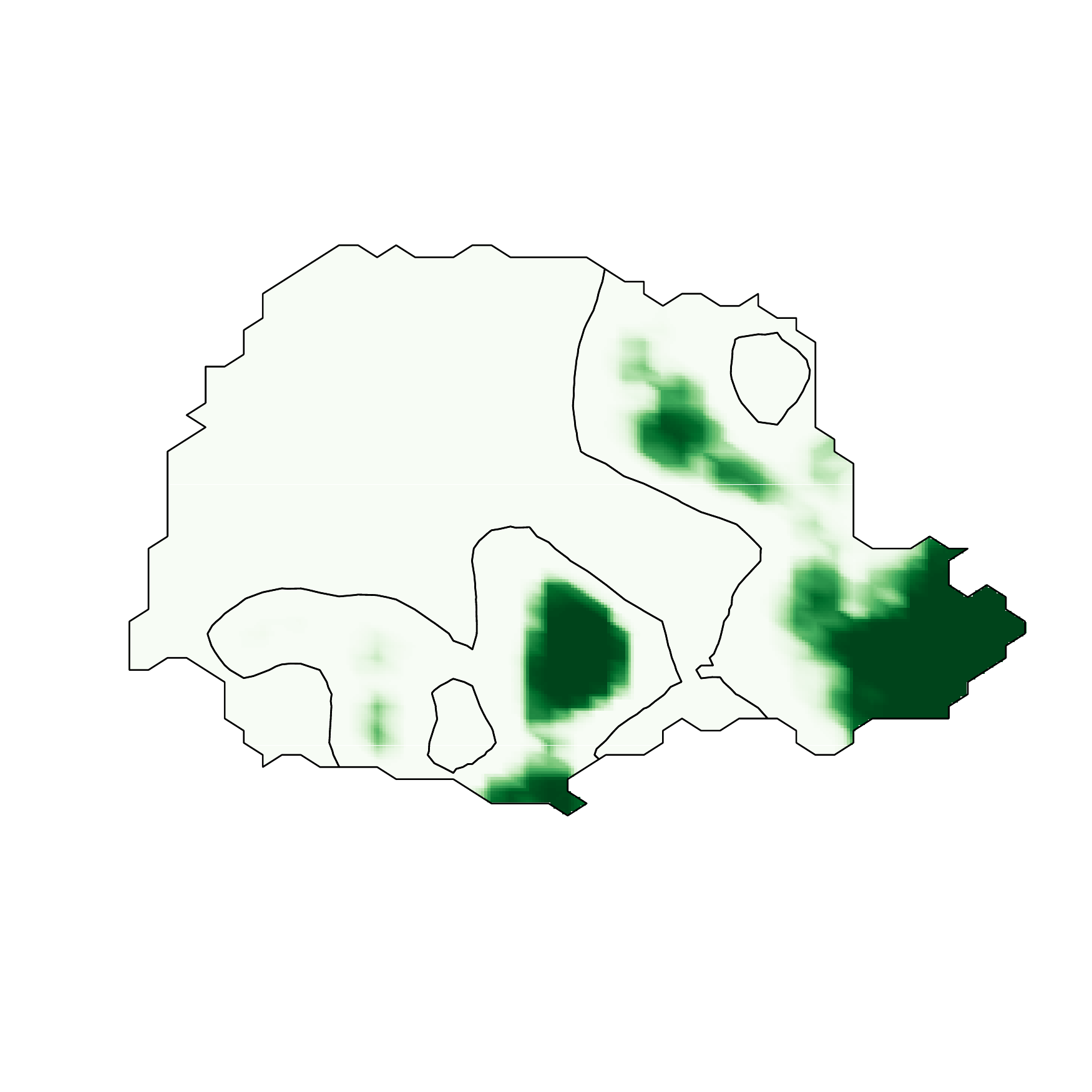}
\raisebox{1cm}{\includegraphics[trim= 434 15 0 0,clip,width=0.05\linewidth]{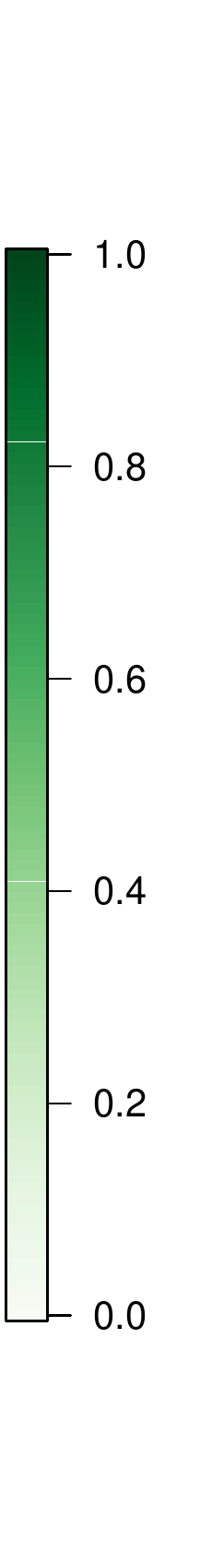}}}
\vspace{-1cm}
\caption{A contour map of the posterior mean of log precipitation (left), and the excursion function for the positive $u=\log(7)$ excursion set (right). In the right panel, also the $u=\log(7)$ contour is shown.}
\label{fig:precip1}
\end{figure}

We now want to find areas that likely experienced large amounts of precipitation. In the following code, we compute the excursion set for the posterior mean for the level $7$ mm of precipitation. To indicate that this level is in the scale of the data, and not in the scale of the linear predictor, we use the \code{u.link=TRUE} argument in the \code{excursions} call.
\begin{Schunk}
\begin{Sinput}
R> exc = excursions.inla(r, stk, tag = "prd", u = 7, u.link = TRUE,
+   type = ">", F.lim = 0.6, method = 'QC')
R> sets <- continuous(exc, submesh, alpha = 0.1)
\end{Sinput}
\end{Schunk}
We also compute the contour curve for the level of interest on the continuous domain, using the \code{tricontourmap} function.
\begin{Schunk}
\begin{Sinput}
R> con <- tricontourmap(submesh, z=exc$mean, levels = log(7))
\end{Sinput}
\end{Schunk}
We plot the resulting continuous domain excursion function together with the contour curve using the following commands. The result is shown in the right panel of Figure~\ref{fig:precip1}.

\begin{Schunk}
\begin{Sinput}
R> proj <- inla.mesh.projector(sets$F.geometry, dims = c(300,200))
R> image(proj$x, proj$y, inla.mesh.project(proj, field = sets$F),
+   col = cmap.F, axes = FALSE, xlab = "", ylab = "", asp = 1)
R> plot(con$map, add = TRUE)
\end{Sinput}
\end{Schunk}

To visualize the posterior mean using a contour map using the following commands
\begin{Schunk}
\begin{Sinput}
R> lp <- contourmap.inla(r, stack = stk, tag = "prd", n.levels = 2,
+   compute = list(F = FALSE))
R> tmap <- tricontourmap(submesh, z = exc$mean, levels = lp$meta$levels)
R> plot(tmap$map, col = contourmap.colors(lp, col = cmap))
\end{Sinput}
\end{Schunk}
Here, the \code{contourmap.inla} computes the levels of the contour map and \code{tricontourmap} computes the contour map on the mesh. Finally, \code{contourmap.colors} is used to compute appropriate colours for visualising the contour map. The result is shown in the left panel of Figure~\ref{fig:precip1}.

The contour map we computed had two contours, and a relevant question is now if this is an appropriate number. To investigate this, we compute the $P_2$ quality measure for this contour map and for contour maps with one and three levels.
\begin{Schunk}
\begin{Sinput}
R> lp1 <- contourmap.inla(r, stack = stk, tag = "prd", n.levels = 1,
+   compute = list(F = FALSE, measures = c("P2")))
R> lp2 <- contourmap.inla(r, stack = stk, tag = "prd", n.levels = 2,
+   compute = list(F = FALSE, measures = c("P2")))
R> lp3 <- contourmap.inla(r, stack = stk, tag = "prd", n.levels = 3,
+   compute = list(F = FALSE, measures = c("P2")))
R> cat(c(lp1$P2, lp2$P2, lp3$P2))
\end{Sinput}
\begin{Soutput}
1 0.4785692 0.004764646
\end{Soutput}
\end{Schunk}
We see that using only one contour gives very high credibility, whereas using three contours give a credibility that is close to zero. The contour map with two contours has a credibility around $0.5$ and seems to be a good choice for this application.

\section{Discussion}\label{sec:discussion}
The \pkg{excursions} package was first developed for calculating probabilistic excursion sets and contour credible regions for latent Gaussian stochastic processes and fields. Since the early versions, the scope of the package has grown and it now contains functions for several other related computations, such as simultaneous confidence bands, uncertainty quantification of contour maps, and computations of Gaussian integrals.

Some of the functionaliy in \pkg{excursions} can also be found in other packages. For example, \pkg{ExceedanceTools} \citep{french2014exceedance} provides an alternative method for computing excursion sets and uncertainty regions for contour curves for Gaussian models. The package \pkg{mvtnorm} \citep{genz2016mvtnorm} contains functions for computing integrals of Gaussian distributions, but does not have support for computations based on sparse precision matrices. Finally, there are numerous packages with the ability to compute simultaneous confidence bands for various models, such as semiparametric regression models using \pkg{AdaptFitOS} \citep{wiesenfarth2012simultaneous} or nonparametric regression models for functional data using \pkg{SCBmeanfd}  \citep{SCBmeanfd}. Comparing the methods in \pkg{excursions} to the corresponding methods in these packages is an intersting topic for future studies.

Future work also includes further development of the package, and new features are added as they are needed. One main focus for the current development is to add functionality for large scale computations, where sparse Cholesky factorisation is computationally infeasible. We also plan to add functionality for computations needed in spatial extreme value theory, where certain Gaussian integrals often are needed during likelihood inference. Finally, the technical aspects of the functions are also being improved, and future releases will introduce improvements to both the continuous domain interpretations and to the algorithms for finding the largest excursion sets.

\bibliography{journ_abrv,completebib}

\end{document}